\documentclass{ws-procs975x65}

\def\beq{\begin{equation}}
\def\eeq{\end{equation}}

\begin{document}

\title{Non-relativistic expansion of the Einstein--Hilbert Lagrangian}
\author{Dennis Hansen$^1$, Jelle Hartong$^{2*}$ and Niels A. Obers$^{3,4}$}

\address{$^1$ Institut f{\"u}r Theoretische Physik, Eidgen{\"o}ssische Technische Hochschule Z{\"u}rich\\ 
Wolfgang-Pauli-Strasse 27, 8093 Z{\"u}rich, Switzerland\\
$^2$ School of Mathematics and Maxwell Institute for Mathematical Sciences,\\
University of Edinburgh, Peter Guthrie Tait road, Edinburgh EH9 3FD, UK\\
$^3$ Nordita, KTH Royal Institute of Technology and Stockholm University,\\
Roslagstullsbacken 23, SE-106 91 Stockholm, Sweden\\
$^4$ The Niels Bohr Institute, Copenhagen University,\\
Blegdamsvej 17, DK-2100 Copenhagen \O , Denmark\\
$^*$E-mail: j.hartong@ed.ac.uk}

%


\begin{abstract}
We present a systematic technique to expand the Einstein--Hilbert Lagrangian in inverse powers of the speed of light squared. The corresponding result for the non-relativistic gravity Lagrangian is given up to next-to-next-to-leading order. The techniques are universal and can be used to expand any Lagrangian theory
whose fields are a function of a given parameter.
\end{abstract}


\bodymatter


\section{Introduction}

Recent developments have significantly improved our understanding of non-relativistic geometries such as Newton--Cartan geometry and their role in approximations and limits of relativistic theories such as general relativity, Chern--Simon theories and string theories. We will show how Newton--Cartan geometry and its torsionful generalization can be used to construct an order by order non-relativistic approximation of the Einstein--Hilbert Lagrangian. This will provide us with a non-relativistic gravity Lagrangian whose equations of motion describe Newtonian gravity and generalizations thereof that involve e.g. gravitational time dilation.

\section{Expanding fields and Lagrangians}\label{sec:1}

Consider a Lagrangian for a set of fields $\phi^I$ (with spacetime indices suppressed) that is a function of $c,\phi^I,\partial_\mu\phi^I$ where $\phi^I=\phi^I_{(0)}+c^{-2}\phi^I_{(1)}+c^{-4}\phi^I_{(2)}+\cdots$ with $c$ the speed of light. The explicit $c$ dependence is determined by the Lagrangian and is related to our choice of working with a set of fields whose $1/c^2$ expansion starts at order $c^0$. We will only consider even powers series in $c^{-1}$.
Assuming the overall power of the Lagrangian is $c^N$ we define $\tilde{\mathcal{L}}(\sigma)=c^{-N}\mathcal{L}(c,\phi^I,\partial_\mu\phi^I)$ where $\sigma=c^{-2}$. We then Taylor expand $\tilde{\mathcal{L}}(\sigma)$ around $\sigma=0$, i.e.
\begin{equation}
\tilde{\mathcal{L}}(\sigma)=\tilde{\mathcal{L}}(0)+\sigma \tilde{\mathcal{L}}'(0)+\frac{1}{2}\sigma^2\tilde{\mathcal{L}}''(0)+O(\sigma^3)\,,
\end{equation}
where the prime denotes differentiation with respect to $\sigma$, which satisfies
\begin{equation}
\frac{d}{d\sigma}=\frac{\partial}{\partial\sigma}+\frac{\partial\phi^I}{\partial\sigma}\frac{\partial}{\partial\phi^I}+\frac{\partial\partial_\mu\phi^I}{\partial\sigma}\frac{\partial}{\partial\partial_\mu\phi^I}\,.
\end{equation}
Hence if we write 
\begin{eqnarray}
\mathcal{L}& = & c^N\tilde{\mathcal{L}}(\sigma)=c^N\left(\mathcal{L}_{\text{LO}}+c^{-2}\mathcal{L}_{\text{NLO}}+c^{-4}\mathcal{L}_{\text{NNLO}}+O(c^{-6})\right)\,,
\end{eqnarray}
it follows that the LO and NLO Lagrangians take the form
\begin{eqnarray}
\mathcal{L}_{\text{LO}} & = & \tilde{\mathcal{L}}(0)=\mathcal{L}_{\text{LO}}(\phi^I_{(0)},\partial_\mu\phi^I_{(0)})\,,\label{eq:expLagrangian1}\\
\mathcal{L}_{\text{NLO}} & = & \tilde{\mathcal{L}}'(0)=\frac{\partial\tilde{\mathcal{L}}}{\partial\sigma}\Big|_{\sigma=0}+\phi^I_{(1)}\frac{\delta\mathcal{L}_{\text{LO}}}{\delta\phi^I_{(0)}}\,.\label{eq:expLagrangian2}
\end{eqnarray}

This shows that the equations of motion of the next-to-leading order (NLO) fields of the next-to-leading order Lagrangian are the equations of motion of the leading order (LO) fields of the leading order Lagrangian. A very similar calculation gives for the NNLO Lagrangian the expression
\begin{eqnarray}
\hspace{-.6cm}&& \mathcal{L}_{\text{NNLO}}  = \frac{1}{2}\frac{\partial^2\tilde{\mathcal{L}}}{\partial\sigma^2}\Big|_{\sigma=0}+\phi^I_{(1)}\frac{\delta}{\delta\phi^I_{(0)}}\frac{\partial\tilde{\mathcal{L}}}{\partial\sigma}\Big|_{\sigma=0}+\phi^I_{(2)}\frac{\delta\mathcal{L}_{\text{LO}}}{\delta\phi^I_{(0)}}+\frac{1}{2}\left[\phi_{(1)}^I\phi_{(1)}^J\frac{\partial^2\mathcal{L}_{\text{LO}}}{\partial\phi_{(0)}^I\partial\phi_{(0)}^J}\right.\nonumber\\
\hspace{-.6cm}&&\left.+2\phi^I_{(1)}\partial_\mu\phi^J_{(1)}\frac{\partial^2\mathcal{L}_{\text{LO}}}{\partial\phi^I_{(0)}\partial(\partial_\mu\phi^J_{(0)})}+\partial_\mu\phi^I_{(1)}\partial_\nu\phi^J_{(1)}\frac{\partial^2\mathcal{L}_{\text{LO}}}{\partial(\partial_\mu\phi^I_{(0)})\partial(\partial_\nu\phi^J_{(0)})}\right]\!\!.\label{eq:NNLO}
\end{eqnarray}
The term in square brackets is the second variation of the LO Lagrangian and is a quadratic form involving the Hessian of the LO Lagrangian. It can be shown that
\begin{equation}
\frac{\delta\mathcal{L}_{\text{NNLO}}}{\delta\phi^I_{(1)}}=\frac{\delta\mathcal{L}_{\text{NLO}}}{\delta\phi^I_{(0)}}\,.
\end{equation}
Combining this with the fact that the equations of motion of $\phi^I_{(2)}$ of the NNLO Lagrangian give the equations of motion of the LO Lagrangian we see that the NNLO Lagrangian reproduces all of the equations of motion of the NLO Lagrangian. 

\section{Explicit dependence on speed of light of Einstein--Hilbert Lagrangian}

In order to apply these ideas  to the expansion of the Einstein--Hilbert (EH) Lagrangian we need to construct fields that start at order $c^0$. To this end we define
\begin{equation}\label{eq:vielbeindecom}
g_{\mu\nu}=-c^2T_\mu T_\nu+\Pi_{\mu\nu}\qquad\text{and}\qquad g^{\mu\nu}=-\frac{1}{c^2}T^\mu T^\nu+\Pi^{\mu\nu}\,,
\end{equation}
where $c^2=\hat c^2/\sigma$ with $\hat c$ equal to the speed of light. The expansion is in $\sigma$ around zero which is the square of the dimensionless slope of the light cone in tangent space. We will set $\hat c=1$ and expand around $1/c^2$. We can always impose the conditions
\begin{equation}
T_\mu\Pi^{\mu\nu}=0\,,\qquad T^\mu\Pi_{\mu\nu}=0\,,\qquad T_\mu T^\mu=-1\,,\qquad\Pi_{\mu\rho}\Pi^{\rho\nu}=\delta^\nu_\mu+T^\nu T_\mu\,.
\end{equation}
It will be convenient to use the following torsionful connection (instead of the usual Levi-Civita connection)
\begin{equation}
C^{\rho}_{\mu\nu} = -T^\rho\partial_\mu T_\nu+\frac{1}{2}\Pi^{\rho\sigma}\left(\partial_\mu\Pi_{\nu\sigma}+\partial_\nu\Pi_{\mu\sigma}-\partial_\sigma\Pi_{\mu\nu}\right)\,.
\end{equation}
The covariant derivative $\overset{\scriptscriptstyle{(C)}}{\nabla}{}_{\!\mu}$  with respect to this connection obeys
\begin{equation}
\overset{\scriptscriptstyle{(C)}}{\nabla}{}_{\!\mu} T_\nu=0\,,\qquad \overset{\scriptscriptstyle{(C)}}{\nabla}{}_{\!\mu}\Pi^{\nu\rho}=0\,,\qquad \overset{\scriptscriptstyle{(C)}}{\nabla}{}_{\!\mu} T^\nu=\frac{1}{2}\Pi^{\nu\rho}\mathcal{L}_T\Pi_{\rho\mu}\,,
\end{equation}
where $\mathcal{L}_T$ is the Lie derivative along $T^\mu$. The EH Lagrangian can be written as 
\begin{eqnarray}
\mathcal{L}_{\text{EH}}=\frac{c^3}{16\pi G}\sqrt{-g}R=\frac{c^6}{16\pi G}\tilde{\mathcal{L}}(\sigma,T,\Pi,\partial)\,,
\end{eqnarray}
where $G$ is Newton's constant. The prefactor $c^3$, as opposed to the usual $c^4$, is due to the presence of $c$ in \eqref{eq:vielbeindecom}. The Lagrangian $\tilde{\mathcal{L}}$ which is at most order $c^0$ is 
\begin{equation}\label{eq:explicitsigma}
\tilde{\mathcal{L}} = \sqrt{-\text{det}\,(-T_\alpha T_\beta+\Pi_{\alpha\beta})}\left[\frac{1}{4}\Pi^{\mu\nu}\Pi^{\rho\sigma}T_{\mu\rho}T_{\nu\sigma}+\sigma\Pi^{\mu\nu}\overset{\scriptscriptstyle{(C)}}{R}{}_{\!\mu\nu}-\sigma^2 T^\mu T^\nu\overset{\scriptscriptstyle{(C)}}{R}{}_{\!\mu\nu}\right].
\end{equation}
In here $\overset{\scriptscriptstyle{(C)}}{R}{}_{\!\mu\nu}$ is defined with respect to the connection $C^\rho_{\mu\nu}$ and $T_{\mu\nu}\equiv\partial_\mu T_\nu-\partial_\nu T_\mu$.

\section{Non-relativistic expansion of the metric}

The fields $T_\mu$ and $\Pi_{\mu\nu}$, entering the metric, admit the following $1/c^2$ expansion
\begin{eqnarray}
\hspace{-.5cm}T_\mu & = & \tau_\mu+c^{-2}m_\mu+c^{-4}B_\mu+O(c^{-6})\,,\qquad\Pi_{\mu\nu} = h_{\mu\nu}+c^{-2}\Phi_{\mu\nu}+O(c^{-4})\,,\\
\hspace{-.5cm}T^\mu & = & v^\mu+O(c^{-2})\,,\qquad\Pi^{\mu\nu}=h^{\mu\nu}+O(c^{-2})\,,
\end{eqnarray}
where the LO fields obey the following orthogonality and completeness relations
\begin{equation}
\tau_\mu v^\mu=-1\,,\qquad\tau_\mu h^{\mu\nu}=0\,,\qquad v^\mu h_{\mu\nu}=0\,,\qquad h^{\mu\rho}h_{\rho\nu}=\delta^\mu_\nu+v^\mu\tau_\nu\,.
\end{equation}
This implies that the metric is expanded as 
\begin{equation}
g_{\mu\nu}=-c^2\tau_\mu\tau_\nu+\bar h_{\mu\nu}+\frac{1}{c^2}\bar\Phi_{\mu\nu}+O(c^{-4})\,,
\end{equation}
where we defined
\begin{equation}\label{eq:metriccomponents}
\bar h_{\mu\nu}=h_{\mu\nu}-2\tau_{(\mu} m_{\nu)}\,,\qquad \bar\Phi_{\mu\nu}=\Phi_{\mu\nu}-m_\mu m_\nu-B_\mu\tau_\nu-B_\nu\tau_\mu\,.
\end{equation}
This is equivalent to metric expansions presented in the literature\cite{Dautcourt:1996pm,VandenBleeken:2017rij}.
In the decomposition \eqref{eq:vielbeindecom} we can perform local Lorentz transformations that act on $T_\mu$ and the vielbeine that form $\Pi_{\mu\nu}$. In the $1/c^2$ expansion these become local Galilean boosts (also known as Milne boosts) at leading order and NLO local boosts at NLO. These leave each order of the metric invariant. The quantities $\tau_\mu$, $\bar h_{\mu\nu}$ and $h^{\mu\nu}$ (the LO term of the inverse metric) are Milne boost invariant. The combination \eqref{eq:metriccomponents} in the metric at order $c^{-2}$ is invariant under both local Galilean and NLO local boosts.

The form of the expansions for $T_\mu$ and $\Pi_{\mu\nu}$ should be preserved in any coordinate system. This means that we must also expand the diffeomorphisms in powers of $1/c^2$. If under a diffeomorphism a tensor $X_{\mu\nu}$ changes as $\delta X_{\mu\nu}=\mathcal{L}_\Xi X_{\mu\nu}$ then we should write $\Xi^\mu=\xi^\mu+c^{-2}\zeta^\mu+O(c^{-4})$. All fields in the expansions are then tensors with respect to diffeomorphisms generated by $\xi^\mu$. The LO fields $\tau_\mu$ and $h_{\mu\nu}$ are invariant under the NLO diffeomorphisms $\zeta^\mu$, while the NLO fields transform according to
\begin{eqnarray}
\delta m_\mu & = & \mathcal{L}_\zeta\tau_\mu=\partial_\mu\Lambda-\Lambda a_\mu+h^{\rho\sigma}\zeta_\sigma(\partial_\rho\tau_\mu-\partial_\mu\tau_\rho)\,,\label{eq:NLOdiffeo1}\\
\delta\Phi_{\mu\nu} & = & \mathcal{L}_{\zeta} h_{\mu\nu}=2\Lambda K_{\mu\nu}+\check\nabla_\mu\zeta_\nu+\check\nabla_\nu\zeta_\mu\,,\label{eq:NLOdiffeo2}
\end{eqnarray}
where $\Lambda=\tau_\mu\zeta^\mu$ and $\zeta_\nu=h_{\nu\mu}\zeta^\mu$ and where $K_{\mu\nu}=-\frac{1}{2}\mathcal{L}_v h_{\mu\nu}$ is the extrinsic curvature and $a_\mu=\mathcal{L}_v\tau_\mu=v^\rho(\partial_\rho\tau_\mu-\partial_\mu\tau_\rho)$ is called the torsion vector. The derivative $\check\nabla_\mu$ is covariant with respect to the torsionful connection
\begin{equation}\label{eq:checkGamma}
\check\Gamma^\rho_{\mu\nu}=C^\rho_{\mu\nu}\big|_{\sigma=0}=-v^\rho\partial_\mu\tau_\nu+\frac{1}{2}h^{\rho\sigma}\left(\partial_\mu h_{\nu\rho}+\partial_\nu h_{\mu\rho}-\partial_\rho h_{\mu\nu}\right)\,.
\end{equation}
The geometry is thus described by the LO fields $\tau_\mu$ and $h_{\mu\nu}$ and the subleading fields $m_\mu$ and $\Phi_{\mu\nu}$ are treated as gauge fields on this non-relativistic geometry. This setup is called type II Newton--Cartan geometry\cite{Hansen:2018ofj}.

\section{Non-relativistic expansion of the Einstein--Hilbert Lagrangian}

We are now in a position to apply the results of section \ref{sec:1} to the case of the EH Lagrangian. Using \eqref{eq:explicitsigma} equation \eqref{eq:expLagrangian1} becomes
\begin{equation}
\mathcal{L}_{\text{LO}}=\frac{e}{4}h^{\mu\nu}h^{\rho\sigma}\tau_{\mu\rho}\tau_{\nu\sigma}\,,
\end{equation}
where we defined $\tau_{\mu\nu}=\partial_\mu\tau_\nu-\partial_\nu\tau_\mu$ and $e=(-\text{det}\,(-\tau_\mu\tau_\nu+h_{\mu\nu}))^{1/2}$.

We can consider $\tau_\mu$, $h_{\mu\nu}$, $m_\mu$ and $\Phi_{\mu\nu}$ to be an independent set of quantities that can be varied independently. The total variation of $\mathcal{L}_{\text{LO}}$ is then given by
\begin{eqnarray}
\hspace{-.5cm}\delta\mathcal{L}_{\text{LO}} & = &  e\left[\frac{1}{8}h^{\mu\nu}h^{\rho\sigma}\tau_{\mu\rho}\tau_{\nu\sigma}h^{\alpha\beta}-\frac{1}{2}h^{\mu\alpha}h^{\nu\beta}h^{\rho\sigma}\tau_{\mu\rho}\tau_{\nu\sigma}\right]\delta h_{\alpha\beta}\nonumber\\
\hspace{-.5cm}&&+e\left[-\frac{1}{4}h^{\mu\nu}h^{\rho\sigma}\tau_{\mu\rho}\tau_{\nu\sigma} v^\alpha+h^{\rho\sigma}a_\rho h^{\nu\alpha}\tau_{\nu\sigma}-e^{-1}\partial_\nu\left(e h^{\mu\nu}h^{\rho\alpha}\tau_{\mu\rho}\right)\right]\delta\tau_\alpha\,.\label{eq:totvarLO}
\end{eqnarray}
We used here that $\delta h^{\mu\nu}=\left(v^\mu h^{\nu\rho}+v^\nu h^{\mu\rho}\right)\delta\tau_\rho-h^{\mu\rho}h^{\nu\sigma}\delta h_{\rho\sigma}$. Contracting the $\tau_\mu$ equation of motion with $\tau_\mu$ tells us that $h^{\mu\nu}h^{\rho\sigma}\tau_{\mu\rho}\tau_{\nu\sigma}=0$. This is a sum of squares and so it implies the vanishing of the twist tensor $h^{\mu\nu}h^{\rho\sigma}\tau_{\mu\rho}=0$ and thus that $\tau\wedge d\tau=0$. This is a causality condition for a non-relativistic spacetime. The spacetime admits a global foliation given by a nowhere vanishing hypersurface orthogonal (clock) 1-form $\tau$. This is known as twistless torsional Newton--Cartan (TTNC) geometry\cite{Christensen:2013lma}. 

From equation \eqref{eq:explicitsigma} we find that
\begin{eqnarray}
\frac{\partial\tilde{\mathcal{L}}}{\partial\sigma}\Big|_{\sigma=0} & = & \sqrt{-\text{det}\,(-T_\alpha T_\beta+\Pi_{\alpha\beta})}\Pi^{\mu\nu}\overset{\scriptscriptstyle{(C)}}{R}{}_{\mu\nu}\big|_{\sigma=0}=eh^{\mu\nu}\check R_{\mu\nu}\,.
\end{eqnarray}
The NLO Lagrangian is of the form \eqref{eq:expLagrangian2} and can thus be written as 
\begin{equation}\label{eq:NLOLag}
\mathcal{L}_{\text{NLO}} =eh^{\mu\nu}\check R_{\mu\nu}+\frac{\delta\mathcal{L}_{\text{LO}}}{\delta\tau_\mu}m_\mu+\frac{\delta\mathcal{L}_{\text{LO}}}{\delta h_{\mu\nu}}\Phi_{\mu\nu}\,,
\end{equation}
where the Ricci tensor is defined with respect to the connection \eqref{eq:checkGamma} as
\begin{eqnarray}
\check{R}_{\mu\nu}=\check {R}_{\mu\rho\nu}{}^{\rho}=-\partial_{\mu}\check{\Gamma}_{\rho\nu}^{\rho}+\partial_{\rho}\check{\Gamma}_{\mu\nu}^{\rho}-\check{\Gamma}_{\mu\lambda}^{\rho}\check{\Gamma}_{\rho\nu}^{\lambda}+\check{\Gamma}_{\rho\lambda}^{\rho}\check{\Gamma}_{\mu\nu}^{\lambda}\,.
\end{eqnarray}

In section \ref{sec:1} we have shown that the equations of motion of the NNLO Lagrangian include all the equations of motion of the NLO Lagrangian. Furthermore, as can be seen from \eqref{eq:NNLO}, the equations of motion of the NNLO fields will reproduce the equations of motion of the LO Lagrangian, i.e.  they will impose the TTNC condition. If we are only interested in those equations of motion that involve at most NLO but not NNLO fields then we can ignore those variations of $\tau_\mu$ that lead to equations of motion for the NNLO fields. Using the TTNC condition on shell it can be seen that the only $\tau_\mu$  variation that does not involve NNLO fields is $\tau_\mu\frac{\delta}{\delta\tau_\mu}$. As far as this variation and those of the other fields are concerned we might as well implement the TTNC condition off shell. We thus conclude that an action for the LO and NLO fields can be obtained by computing the NNLO Lagrangian with TTNC off shell. The TTNC condition itself can always be obtained by adding a Lagrange multiplier of the form $B_\mu\frac{\delta\mathcal{L}_{\text{LO}}}{\delta\tau_\mu}$ where $B_\mu$ is a NNLO field. 


We will call the NNLO Lagrangian with TTNC imposed off shell via a Lagrange multiplier the non-relativistic gravity (NRG) Lagrangian. Using \eqref{eq:NNLO} it can be shown to take the form
\begin{eqnarray}
\mathcal{L} & = & e\left[-v^\mu v^\nu\check R_{\mu\nu}-2m_\nu\check\nabla_\mu\left(h^{\mu\rho}h^{\nu\sigma}-h^{\mu\nu}h^{\rho\sigma}\right)K_{\rho\sigma}+\Phi h^{\mu\nu}\check R_{\mu\nu}+\frac{1}{4}h^{\mu\rho}h^{\nu\sigma}F_{\mu\nu}F_{\rho\sigma}\right.\nonumber\\
&&\hspace{-.5cm}\left.-\Phi_{\rho\sigma}h^{\mu\rho}h^{\nu\sigma}\left(\check R_{\mu\nu}-\check\nabla_\mu a_\nu-a_\mu a_\nu-\frac{1}{2}h_{\mu\nu}h^{\kappa\lambda}\check R_{\kappa\lambda}+h_{\mu\nu}e^{-1}\partial_\kappa\left(eh^{\kappa\lambda}a_\lambda\right)\right)\right]\,,
\label{eq:action3}
\end{eqnarray}
where $\Phi=-v^\mu m_\mu$ is the Newtonian potential and where $F_{\mu\nu}=\partial_\mu  m_\nu-\partial_\nu m_\mu-a_\mu m_\nu+a_\nu m_\mu$. We did not write the Lagrange multiplier term.
When comparing with \eqref{eq:NNLO} we see that the first term in \eqref{eq:action3} agrees with $\frac{1}{2}\frac{\partial^2\tilde{\mathcal{L}}}{\partial\sigma^2}\big|_{\sigma=0}$. The terms linear in $\Phi_{\mu\nu}$ and $m_\mu$ agree with $\phi^I_{(1)}\frac{\delta}{\delta\phi^I_{(0)}}\frac{\partial\tilde{\mathcal{L}}}{\partial\sigma}\big|_{\sigma=0}$. The terms containing NNLO fields in \eqref{eq:NNLO} are not present here because we used TTNC off shell. Finally the second order variation of the LO Lagrangian with respect to $\tau_\mu$ is still nontrivial even when we use TTNC off shell and this leads to the field strength squared terms in \eqref{eq:action3} on the last line. Note that there are no terms quadratic in $\Phi_{\mu\nu}$ because all derivatives with respect to $h_{\mu\nu}$ of the LO Lagrangian give zero upon using the TTNC condition. The equations of motion of $\Phi_{\mu\nu}$ and $m_\mu$ agree with the equations of motion of $h_{\mu\nu}$ and $\tau_\mu$ when varying \eqref{eq:NLOLag}.

There is only one term containing $\Phi=-v^\mu m_\mu$, the Newtonian potential. To find Newtonian gravity and its generalization to TTNC geometries requires computing the variations $\tau_\mu\frac{\delta}{\delta\tau_\mu}$ and $h_{\mu\nu}\frac{\delta}{\delta h_{\mu\nu}}$\cite{Hansen:2018ofj}. In order to continue expanding to higher orders we need to compute the full NNLO Lagrangian without using TTNC off shell. 

In non-relativistic gravity $\tau$ is of the form $\tau=NdT$ where $N$ is a non-relativistic lapse function and $T$ is some time function. This means that gravitational time dilation is a non-relativistic phenomenon\cite{VandenBleeken:2017rij,Hansen:2019vqf}. For example the Tolman--Oppenheimer--Volkoff solution for a fluid star is a solution of NRG coupled to a fluid\cite{VandenBleeken:2019gqa,Hansen:2020pqs}. As a result the classical tests of general relativity: gravitational redshift, perihelion precession and bending of light are also passed by non-relativistic gravity\cite{Hansen:2019vqf}.

We can rewrite \eqref{eq:action3} in terms of manifest Milne boost invariant tensors as
\begin{eqnarray}\label{eq:action2}
\hspace{-.5cm}&&\mathcal{L} = e\left[-\hat v^\mu\hat v^\nu\bar R_{\mu\nu}+\hat\Phi h^{\mu\nu}\bar R_{\mu\nu}-\bar\Phi_{\rho\sigma}h^{\mu\rho}h^{\nu\sigma}\Big(\bar R_{\mu\nu}-\bar\nabla_\mu a_\nu-a_\mu a_\nu\right.\nonumber\\
\hspace{-.5cm}&&\left.-\frac{1}{2}h_{\mu\nu}h^{\kappa\lambda}\bar R_{\kappa\lambda}+h_{\mu\nu}e^{-1}\partial_\kappa\left(eh^{\kappa\lambda}a_\lambda\right)\Big)\right]\,.
\end{eqnarray}
where $\hat v^\mu=v^\mu-h^{\mu\nu}m_\nu$ and $\hat\Phi =-v^\mu m_\mu+\frac{1}{2}h^{\mu\nu}m_\mu m_\nu$. In here the Ricci tensors are defined with respect to the Milne boost invariant connection 
\begin{equation}
\bar\Gamma^\rho_{\mu\nu}=-\hat v^\rho\partial_\mu\tau_\nu+\frac{1}{2}h^{\rho\sigma}\left(\partial_\mu\bar h_{\nu\rho}+\partial_\nu\bar h_{\mu\rho}-\partial_\rho\bar h_{\mu\nu}\right)\,.
\end{equation}
The NRG Lagrangian is uniquely fixed by all its gauge symmetries\cite{Hansen:2018ofj}.

It would be interesting to make contact with the post-Newtonian approximation by finding a convenient way of expressing the higher orders of the $1/c^2$ expansion of the EH Lagrangian and more generally to see if this off shell and covariant approach can be used for interesting approximations to general relativity. To this end it might be useful to employ a first order formulation\cite{Cariglia:2018hyr,Bergshoeff:2019ctr}.

\section*{Acknowledgments}

We thank Dieter Van den Bleeken for useful discussions.
The work of DH is supported by the Swiss National Science Foundation through the NCCR SwissMAP.
The work of JH is supported by the Royal Society University Research Fellowship ``'Non-Lorentzian Geometry in Holography'' (UF160197).
The work of NO is supported in part by the project ``Towards a deeper understanding of black holes with non-relativistic holography'' of the Independent Research Fund Denmark (DFF-6108-00340) and Villum Foundation Experiment project 00023086.

\end{document}